\documentclass[prc,twocolumn,superscriptaddress,superscriptfootnote]{revtex4}
\usepackage{amssymb}

\usepackage{amsmath,amsfonts,amssymb,epsfig,color}

\begin{document}
\title{Triaxial Shapes in the Interacting Vector
Boson Model}
\author{H. G. Ganev}
\affiliation{Institute of Nuclear Research and Nuclear Energy,
Bulgarian Academy of Sciences, \\ Sofia 1784, Bulgaria\\}
\affiliation{Bogoliubov Laboratory of Theoretical Physics, Joint
Institute for Nuclear Research, \\
141980 Dubna, Moscow Region, Russia}

\setcounter{MaxMatrixCols}{10}

\begin{abstract}

A new dynamical symmetry limit of the two-fluid Interacting Vector
Boson Model (IVBM), defined through the chain $Sp(12,R) \supset
U(3,3) \supset U^{\ast}(3) \otimes SU(1,1) \supset SU^{\ast}(3)
\supset SO(3)$, is introduced. The $SU^{\ast}(3)$ algebra considered
in the present paper closely resembles many properties of the
$SU^{\ast}(3)$ limit of IBM-2, which have been shown by many authors
geometrically to correspond to the rigid triaxial model. The
influence of different types of perturbations on the $SU^{\ast}(3)$
energy surface, in particular the addition of a Majorana interaction
and an $O(6)$ term to the model Hamiltonian, is studied. The effect
of these perturbations results in the formation of a stable triaxial
minimum in the energy surface of the IVBM Hamiltonian under
consideration. Using a schematic Hamiltonian which possesses a
perturbed $SU^{\ast}(3)$ dynamical symmetry, the theory is applied
for the calculation of the low-lying energy spectrum of the nucleus
$^{192}$Os. The theoretical results obtained agree reasonably with
the experimental data and show a very shallow triaxial minimum in
the energy surface for the ground state in $^{192}$Os, suggesting
that the newly proposed dynamical symmetry might be appropriate for
the description of the collective properties of different nuclei,
exhibiting triaxial features.

\end{abstract}
\maketitle PACS number(s): {21.60.Ev, 21.60.Fw,}

\section{Introduction}

It has been known for a long time that in certain mass regions
nuclei with static deformation show deviations from a rigid axially
symmetric picture. The possibility of static triaxial shapes for the
ground state of nuclei is a long-standing problem in nuclear
structure physics despite the fact that very few candidates have
been found experimentally \cite{exp1},\cite{exp2}. In the
geometrical approach the triaxial nuclear properties are usually
interpreted in terms of either the $\gamma-$unstable (or
$\gamma-$soft) rotor model of Wilets and Jean \cite{WJ} or the rigid
triaxial rotor model of Davydov \emph{et al.} \cite{DF}. These
models exploit the geometrical picture of nucleus according to the
Collective Model (CM) of Bohr and Mottelson, expressed  in terms of
the intrinsic variables $\beta$ and $\gamma$ where the former
specifies the ellipsoidal quadrupole deformation and the latter the
degree of axial asymmetry. To describe the deviations from axial
symmetry the model of Wilets and Jean assumes that the potential
energy is independent of the $\gamma$-degree of freedom, while in
the model of Davydov \emph{et al.} one considers a harmonic
oscillator potential with a minimum at finite values of $\gamma$
producing a rigid triaxial shape of the nucleus.

Recently, analytical solutions of the Bohr Hamiltonian regarding the
triaxial shapes using a Davidson potential \cite{Z5D} and a sextic
oscillator \cite{SP} have been obtained, where the triaxial shapes
are assumed from the very beginning. The former, called $Z(5)-D$
solution, is shown to cover the region between a triaxial vibrator
and the rigid triaxial rotator, while the $Z(5)$ solution
corresponds to the critical point of the shape phase transition from
a triaxial vibrator to the rigid triaxial rotator. Triaxiality has
been also studied in the framework of the algebraic collective model
\cite{ACM1}, and the onset of rigid triaxial deformation has been
considered \cite{ACM2}.

An alternative description of nuclear collective excitations is
provided by the IBM, which in contrast to the geometrical models, is
of an algebraic nature. To accommodate the triaxial shapes in the
IBM, several approaches can be adopted. It was shown that the
triaxial shapes can occur in three different cases:

(i) In the IBM-1 framework, in which no distinction between protons
and neutrons is made, the inclusion of higher-order (three-body)
terms is needed \cite{cube1},\cite{cube2}.

(ii) In the sdg-IBM framework (using s, d, and g bosons), the
presence of the $g$ boson also suffices \cite{gbos1}, \cite{sdgIBM}.

(iii) In the IBM-2 framework, in which protons and neutrons are used
as distinct entities, the inclusion of one-body and two-body terms
suffices \cite{PSDiep},\cite{PSIBM2a},\cite{PSIBM2b},\cite{IBMQPT}.

In the IBM-1 framework the triaxial shapes are usually obtained by
adding three-body terms of the type
$[d^{\dag}d^{\dag}d^{\dag}]^{(L)} \cdot
[\widetilde{d}\widetilde{d}\widetilde{d}]^{(L)}$ (see, e.g. Ref.
\cite{cube1}). These terms generate a relatively broad region of
triaxiality in the parameter space of the Hamiltonian. In Ref.
\cite{CCQH} it is shown that by adding to the Consistent-Q Formalism
Hamiltonian a cubic combination $(\widehat{Q} \times \widehat{Q}
\times \widehat{Q})^{(0)}$ of the most general quadrupole operator
$\widehat{Q}^{\chi}$ coupled to zero angular momentum, called the
Cubic Consistent-Q Hamiltonian, there exist a very tiny region of
triaxiality around $\chi \approx \pm \sqrt{7}/2$ between the prolate
and oblate phases. The $\widehat{Q}$ cubic term is interpreted as a
correction to the quadrupole-quadrupole scalar product, which  in
combination with the latter can generate stable triaxial shapes.

The study of the effects of various multipole interactions within
the framework of the sdg-IBM on the equilibrium shape of the ground
state in deformed nuclei has revealed that a hexadecapole
interaction involving a g-boson is needed in order to induce a
triaxial shape. Over the years, both microscopic and
phenomenological evidence has been gathered that shows the
importance of the g-boson in deformed regions. Including the $g$
boson, in a recent study in the sdg-IBM no shape or phase
transitions toward stable triaxial shapes was found \cite{sdgIBM}.

An important feature offered by the IBM-2 \cite{IBM2} is the
possibility to get triaxial shapes
\cite{PSDiep},\cite{PSIBM2a},\cite{PSIBM2b},\cite{IBMQPT} besides
the axially symmetric ones only taking into account explicitly the
proton-neutron degrees of freedom or the two-fluid character of the
nuclear system. The triaxial shapes then arise as a result of
different deformations of the proton and neutron fluids. The
microscopic conditions leading to two-fluid triaxial structure are
found when the proton bosons are particle like (i.e. below
mid-shell) and the neutron bosons are hole-like (above mid-shell) or
vice versa. A new critical point $Y(5)$ symmetry \cite{Y5} from
axially deformed to triaxial shapes was proposed and suggested to be
of importance of considering triaxial shapes in the phase diagram of
the IBM-2.

In the present paper we exploit an algebraic approach, complementary
to IBM, for the description of triaxial nuclei and show how within
the framework of the phenomenological Interacting Vector Boson Model
(IVBM) one might obtain triaxial shapes. The IVBM and its recent
applications for the description of diverse collective phenomena in
the low-lying energy spectra (see, e.g., the review article
\cite{pepan}) exploit the symplectic algebraic structures and the
Sp(12,$R$) is used as a dynamical symmetry group. Symplectic
algebras have been applied extensively in the theory of nuclear
structure. They are used generally to describe systems with a
changing number of particles or excitation quanta and in this way
provide for larger representation spaces and richer subalgebraic
structures that can accommodate the more complex structural effects
as realized in nuclei with nucleon numbers that lie far from the
magic numbers of closed shells.

The symplectic symmetries emerge as appropriate dynamical symmetries
for the many-body theory of collective motion, considering the
nucleus from a hydrodynamic perspective \cite{str}. For example, the
one-fluid symplectic model of Rowe and Rosensteel \cite{SMRR}, based
on the non-compact dynamical algebra $Sp(6,R)$, allows for the
description of rotational dynamics in a continuous range from
irrotational to rigid rotor flows. The extension of the $Sp(6,R)$
symplectic model to the case of two-fluid nuclear systems leads
naturally to the $Sp(12,R)$ dynamical symmetry. In this respect the
symplectic IVBM can be considered as a generalization of the
symplectic model of Rowe and Rosensteel (contained as a submodel of
the $Sp(12,R)$ IVBM), when the nuclear many-body system is viewed as
consisting of two different interacting subsystems.

The different shapes that take place within the framework of the
two-fluid IVBM have been investigated in Ref. \cite{PSIVBM}. It has
been shown that there exist three distinct shapes corresponding to
the three dynamical symmetries of IVBM:\quad (1) spherical shape,
$U_{p}(3)\otimes U_{n}(3)$, \quad (2) $\gamma-$unstable deformed
shape, $O(6)$, and \quad (3) axially deformed shape, $SU(3)\otimes
U_{T}(2)$. It turns out that these are not all possible shapes
associated with the algebraic structures of the IVBM that might
arise. The aim of this paper is to show that the IVBM possesses a
very rich phase structure, which also contains, beyond the spherical
and axially deformed shapes, triaxial shapes. For this purpose we
propose a new dynamical symmetry limit of the IVBM, which in some
aspects is related to the one of the dynamical symmetries of the
IBM-2, namely the $SU^{\ast}(3)$ one. The $SU^{\ast}(3)$ limit of
IBM-2 has been discussed extensively in Refs.
\cite{SU3AST1},\cite{SU3AST2},\cite{SU3AST3}. The latter gives rise
to the Dieperink tetrahedron \cite{PSDiep}, which has an extra
dimension compared to the Casten triangle \cite{triangle}, and to a
new, triaxial shape phase of the model.

It has been shown in the literature that the exact $SU^{\ast}(3)$
symmetry possesses a large degeneracy in the level spectra which in
actual nuclei is not observed and hence the $SU^{\ast}(3)$ symmetry
probably does not appear in its pure form and must be perturbed. In
many cases, the energy spectra exhibit transitional patterns and
might be situated in between the $SU^{\ast}(3)$ and $O(6)$ or
$SU(3)$ and $SU^{\ast}(3)$ dynamical limits. In this respect, we
study the influence of different types of perturbations on the
$SU^{\ast}(3)$ dynamical symmetry energy surface of the IVBM. It is
shown that the newly proposed dynamical symmetry limit might be of
relevance for the description of the collective properties of
different nuclei exhibiting triaxial features.

\section{The algebraic structure of the new dynamical symmetry}

It was suggested by Bargmann and Moshinsky \cite{BargMosh} that two
types of bosons are needed for the description of nuclear dynamics.
It was shown there that the consideration of only two-body system
consisting of two different interacting vector particles will
suffice to give a complete description of $N$ three-dimensional
oscillators with a quadrupole-quadrupole interaction. The latter can
be considered as the underlying basis in the algebraic construction
of the \emph{phenomenological} IVBM.

The algebraic structure of the IVBM
\cite{PSIVBM},\cite{IVBM},\cite{pepan} is realized in terms of
creation and annihilation operators of two kinds of vector bosons
$u_{m}^{\dag}(\alpha )$, $u_{m}(\alpha )$ ($m=0,\pm 1$), which
differ in an additional quantum number $\alpha=\pm1/2$ (or
$\alpha=p$ and $n$)$-$the projection of the $T-$spin (an analogue to
the $F-$spin of IBM-2 or the $I-$spin of the particle-hole IBM). In
the present paper, we consider these two bosons just as elementary
building blocks or quanta of elementary excitations (phonons) rather
than real fermion pairs, which generate a given type of algebraic
structures. Thus, only their tensorial structure is of importance
and they are used as an auxiliary tool, generating an appropriate
\emph{dynamical} symmetry. These elementary excitations carry an
angular momentum $l=1$, i.e. they transform as vectors with respect
the rotational group $SO(3)$. In this regard, the $s$ and $d$ bosons
of the IBM-1 can be considered as bound states of elementary
excitations generated by the two vector bosons.

The microscopic foundation of the IVBM is beyond the scope of the
present paper. (A short discussion on this matter can be found in
Ref.\cite{pepan}.) Nevertheless, some remarks concerning this topic
can be very useful for the readers who are not familiar with the
IVBM.

It is known that the IBM is now standard and the $s$ and $d$ bosons
are viewed as working approximations of the composite $S$ and $D$
bosons made up of nucleons held together by the pairing and the
quadrupole forces. Additional degrees of freedom are further
incorporated in the extended versions of the model (e.g. the
inclusion of $p$, $f$ and $g$ bosons; the inclusion of the isospin,
the F-spin and the particle-hole I-spin). In this respect, the
natural question about the connection between the IVBM and the
standard versions of IBM arises. The answer is obtained
\cite{Asherova} by means of the boson mapping technique, which is
widely applied to the problems of microscopic foundation of IBM
\cite{MFIBM}. It is shown \cite{Asherova} that the IVBM boson space
can be mapped on the ideal boson space of IBM including beyond the
standard $s$ and $d$ bosons (IBM-1), also the $p$ bosons. The latter
(together with the $f$ bosons) are shown to play a crucial role in
the description of the deformed asymmetric shapes in nuclei, in
which the octupole and dipole (cluster) degrees of freedom must be
taken into account. This specific version of IBM is denoted as
IBM-3.5 (intermediate between IBM-3 and IBM-4). The interaction
between these secondary $s$, $d$ and $p$ bosons is induced by the
interaction between the vector bosons.

A similar situation occurs also in the specific isospin-invariant
version of the Fermion Dynamical Dymmetry Model \cite{IFDSM} applied
to the sd-shell nuclei, in which the states constructed from the
nucleon pairs are built from two p-objects ($l =1$), as well as in
the IVBM.

The introduction of a p-boson (p-object) in nuclei with mixed
quadrupole-octupole deformation has been pointed out by many
authors, including also microscopic considerations \cite{Otsuka},
\cite{EI}. The need for the p-boson has been suggested by schematic
shell-model calculations \cite{Catara}, in which collective pairs of
both positive ($S$- and $D$-pairs) and negative parity ($P$- and
$F$-pairs) are used as building blocks. The $p$-boson has been
introduced in different studies of clustering phenomena in nuclei as
well, where the dipole degrees of freedom are connected with the
relative motion of the clusters \cite{cluster}.

In the most general case the two-body model Hamiltonian should be
expressed in terms of the generators of the group $Sp(12,R)$. In
addition to the non-compact "symplectic dynamical symmetry limits"
(subgroup chains starting with some of the symplectic subalgebras of
$Sp(12,R)$; see Refs.\cite{pepan},\cite{SDS}), in some special cases
the two-body model Hamiltonian can be written in terms of the
generators of the subgroups of the maximal compact subgroup $U(6)
\subset Sp(12,R)$ only. The following lattice of group-subgoup
chains of $Sp(12,R)$ takes place (excluding the "symplectic limits",
given in \cite{pepan},\cite{SDS}):
\begin{widetext}
\begin{equation}
\begin{tabular}{lllllllll}
&  &  &  & $U(3)\otimes U_{T}(2)$ & $\longrightarrow $ &
$SU(3)\otimes
U_{T}(2)$ &  &  \\
&  &  & $\nearrow $ &  &  &  & $\searrow $ &  \\
&  & $U(6)$ & $\longrightarrow $ & $\ \ \ \ \ \ \ \ O_{\pm }(6)$ & $%
\longrightarrow $ & $\overline{SU_{\pm }(3)}\otimes SO(2)$ & $\searrow $ &  \\
& $\nearrow $ &  & $\searrow $ &  & $\searrow $ &  &  & $SO(3)$ \\
$Sp(12,R)$ &  &  &  & $U_{p}(3)\otimes U_{n}(3)$ & $\longrightarrow $ & $%
SO_{p}(3)\otimes SO_{n}(3)$ & $\nearrow $ &  \\
& $\searrow $ &  &  & $\ \ \ \ \ \ \ \ \ \ \downarrow $ &  &  &
$\nearrow $
&  \\
&  & $U(3,3)$ & $\longrightarrow $ & $SU_{p}(3)\otimes SU_{n}(3)$ & $%
\longrightarrow $ & $\ \ \ \ \ \ \ SU^{\ast }(3)$ &  &  \\
&  &  & $\searrow $ &  &  $\nearrow $ &  &  &  \\
&  &  &  & $U^{\ast }(3)\otimes U(1,1)$ &  &  &  &
\end{tabular}
\label{chains}
\end{equation}
\end{widetext}

Compared to the lattice given in Ref. \cite{PSIVBM}, here a new
reduction chain (the last one in Eq.(\ref{chains})) is considered.
As it can be seen, the IVBM has a very rich algebraic structure of
subgroups. The first three dynamical limits of the IVBM given in
Eq.(\ref{chains}) and the geometries corresponding to them are
considered in \cite{PSIVBM}. In this paper we are concentrating on
the last reduction chain of the dynamical symmetry group $Sp(12,R)$
of the IVBM for studying the triaxiality in atomic nuclei. As we
will see throughout the paper, this dynamical symmetry is
appropriate for nuclei in which the one type of particles is
particle-like and the other is hole-like.

All bilinear operators of the creation and annihilation operators of
the two kinds of vector bosons
\begin{equation}
u_{k}^{\dag}(\alpha)u_{m}^{\dag}(\beta), \ \ \ \
u_{k}^{\dag}(\alpha)u_{m}(\beta), \ \ \ \ \
u_{k}(\alpha)u_{m}(\beta)
 \label{BRSp12R}
\end{equation}
define the boson representation of the $Sp(12,R)$ algebra. We also
introduce the following notations $u_{m}^{\dag}(\alpha=1/2
)=p^{\dag}_{m}$ and $u_{m}^{\dag}(\alpha=-1/2 )=n^{\dag}_{m}$. In
terms of the $p-$ and $n-$boson operators, the Weyl generators of
the ladder representation of $U(3,3)$ are
\begin{equation}
p_{k}^{\dag}p_{m}, \ \ \ \ p_{k}^{\dag}n_{m}^{\dag}, \ \ \ \ \
-n_{k}p_{m}, \ \ \ \ -n_{m}^{\dag}n_{k},
 \label{BRU33}
\end{equation}
which are obviously a subset of symplectic generators
(\ref{BRSp12R}). The first-order Casimir operator of $U(3,3)$ is
\begin{equation}
C_{1}[U(3,3)]=\sum_{k}(p_{k}^{\dag}p_{k}-n_{k}^{\dag}n_{k}),
 \label{FCOU33}
\end{equation}
and does not differ essentially from the operator $T_{0}$ defined in
\cite{PSIVBM}:
\begin{equation}
T_{0}=\frac{1}{2}C_{1}[U(3,3)]+\frac{3}{2}.
\end{equation}
The algebra $U^{\ast}(3)=\{A_{km} \equiv
p_{k}^{\dag}p_{m}-n_{m}^{\dag}n_{k}\}$ can also be defined in the
following way
\begin{align}
&M=N_{p}-N_{n}, \label{M} \\
&L_M = L^{p}_{M}+L^{n}_{M}, \label{LM} \\
&Q_M = Q^{p}_{M}-Q^{n}_{M}, \label{Qminus}
\end{align}
where the one-fluid operators entering in
($\ref{M}$)$-$($\ref{Qminus}$) are given by
\begin{align}
&N_{p}=\sqrt{3}(p^{\dagger}\times p)^{(0)}, \label{Np} \\
&L^{p}_{M}=\sqrt{2}(p^{\dagger}\times p)^{(1)}_{M}, \label{Lp} \\
&Q^{p}_{M}=\sqrt{2}(p^{\dagger}\times p)^{(2)}_{M}. \label{Qp}
\end{align}
and
\begin{align}
&N_{n}=\sqrt{3}(n^{\dagger}\times n)^{(0)}, \label{Nn} \\
&L^{n}_{M}=\sqrt{2}(n^{\dagger}\times n)^{(1)}_{M}, \label{Ln} \\
&Q^{n}_{M}=\sqrt{2}(n^{\dagger}\times n)^{(2)}_{M}. \label{Qn}
\end{align}

The $U(1,1)$ generators can be obtained from the $U(3,3)$ ones
(\ref{BRU33}) simply by contraction. As will be shown later, the two
algebras $U(1,1)$ and $U^{\ast}(3)$ are mutually complimentary
within a given irrep of $U(3,3)$ \cite{Quesne}.

The second order Casimir operator of $U^{\ast}(3)$ can be defined as
\begin{equation}
C_{2}[U^{\ast}(3)]=\sum_{ij}A_{ij}A_{ji}. \label{C2U*3}
\end{equation}
The $SU^{\ast}(3)$ algebra is obtained by excluding the operator
($\ref{M}$) which is the single generator of the $O(2)$ algebra,
whereas the angular momentum algebra $SO(3)$ is generated by the
generators $L_{M}$ only.

The $U(3,3)$ irreps are positive discrete series irreps
characterized by their lowest weight
$\{f_{3}+\frac{1}{2},f_{2}+\frac{1}{2},f_{1}+\frac{1}{2},
f'_{3}+\frac{1}{2},f'_{2}+\frac{1}{2},f'_{1}+\frac{1}{2}\}$, where
$\{f_{1}+\frac{1}{2},f_{2}+\frac{1}{2},f_{3}+\frac{1}{2}\}$ and
$\{f'_{1}+\frac{1}{2},f'_{2}+\frac{1}{2},f'_{3}+\frac{1}{2}\}$ are
two partitions. The lowest-weight state of such irreps is also the
lowest weight state of an irrep
$\{f_{1}+\frac{1}{2},f_{2}+\frac{1}{2},f_{3}+\frac{1}{2}\} \otimes
\{f'_{1}+\frac{1}{2},f'_{2}+\frac{1}{2},f'_{3}+\frac{1}{2}\}$ of the
maximal compact subgroup $U_{p}(3) \otimes U_{n}(3)$. It turns out
that there exist three cases for the partitions \cite{Quesne}: $(i)$
$f_{1}=\nu
> 0$, $f_{2}=f_{3}=f'_{1}=f'_{2}=f'_{3}=0$;
$(ii)$ $f'_{1}=-\nu > 0$, $f_{1}=f_{2}=f_{3}=f'_{2}=f'_{3}=0$ and
$(iii)$ $f_{1}=f_{2}=f_{3}=f'_{1}=f'_{2}=f'_{3}=\nu=0$. The $U(3,3)$
irreps contained in either irrep $<(1/2)^{6}>$ or $<(1/2)^{5}3/2>$
of $Sp(12,R)$ can be denoted by the shorthand notation $[\nu]$, $\nu
\in Z$, defined as follows:
\begin{align}
[\nu]&=\{(1/2)^{2},\nu+\frac{1}{2};(1/2)^{3}\} \ \ \ \ \emph{if}\ \ \nu > 0 \label{nu1} \\
&=\{(1/2)^{3};(1/2)^{2},-\nu+\frac{1}{2}\} \ \ \emph{if}\ \ \nu < 0 \label{nu2} \\
&=\{(1/2)^{3};(1/2)^{3}\}, \ \ \ \ \ \ \ \ \ \ \ \  \emph{if}\ \ \nu
= 0 \label{nu3}
\end{align}
The branching rules can be written as
\begin{equation}
<(1/2)^{6}> \ \ \downarrow \sum_{{\nu=-\infty},{\nu=even}}^{+\infty}
\oplus[\nu] \label{EIR}
\end{equation}
and
\begin{equation}
<(1/2)^{5}3/2> \ \ \downarrow
\sum_{{\nu=-\infty},{\nu=odd}}^{+\infty} \oplus[\nu]. \label{OIR}
\end{equation}
It can be shown \cite{Quesne} that the label $\nu$ specifying the
$U(3,3)$ irreps in Eqs. (\ref{nu1})-(\ref{OIR}) has a very simple
meaning: it is just the eigenvalue of the first order Casimir
operator (\ref{FCOU33}) of $U(3,3)$, i.e., $\nu=N_{p}-N_{n}$.

The $U^{\ast}(3)$ irreps are characterized by their highest weight
$[n_{1},n_{2},n_{3}]_{3}$, where $n_{1},n_{2},n_{3}$ are some
integers satisfying the inequalities $n_{1}\geq n_{2}\geq n_{3}$. We
note that $[n_{1},n_{2},n_{3}]_{3}$ may assume negative as well as
non-negative values and hence correspond to mixed irreps of
$U^{\ast}(3)$ \cite{Flores}.

The $U(1,1)$ irreps contained in a positive series irrep $[\nu]$ of
$U(3,3)$ are also positive discrete series irreps characterized by
their lowest weight ${N_{p}+\frac{3}{2},N_{n}+\frac{3}{2}}$
\cite{Quesne}. We denote such irreps by the shorthand notation
$[N_{p},N_{n}]=\{N_{p}+\frac{3}{2},N_{n}+\frac{3}{2}\}$. The
$U(1,1)$ and $U^{\ast}(3)$ groups are complementary within any irrep
$[\nu]$ of $U(3,3)$ or, in other words, the irreps $[N_{p},N_{n}]
\otimes [n_{1},n_{2},n_{3}]_{3}$ of $U(1,1) \otimes U^{\ast}(3)$,
contained in a given irrep $[\nu]$ of $U(3,3)$, are multiplicity
free and there is a one-to-one correspondence between the labels
$[n_{1},n_{2},n_{3}]_{3}$ of the $U^{\ast}(3)$ irreps and the labels
$[N_{p},N_{n}]$ of the associated $U(1,1)$ irreps. The precise
relation is $[n_{1},n_{2},n_{3}]_{3} \equiv [N_{p},0,-N_{n}]_{3}$
and $N_{p}-N_{n}=\sum_{k=1}^{3}n_{k}=\nu$ \cite{Quesne}. Then the
$SU^{\ast}(3)$ irreps are $(\mu,\nu)=(N_{p},N_{n})$. This is just
the case when the one type of particles is particle-like and the
other is hole-like and the corresponding algebra can be identified
with the $SU^{\ast}(3)$ one defined in
\cite{SU3AST1},\cite{SU3AST2},\cite{SU3AST3}. Indeed, the
$SU^{\ast}(3)$ algebra can be related to the
$SU(3)=\{L_{M}=L^{p}_{M}+L^{n}_{M},Q_{M}=Q^{p}_{M}+Q^{n}_{M}\}$ one
defined in \cite{PSIVBM} by means of the transformation
\begin{align}
&n^{\dag}_{k} \rightarrow n_{k}, \notag \\
&n_{k} \rightarrow - n^{\dag}_{k},  \label{Aut}
\end{align}
which actually coincides with the particle-hole conjugation.
According to this the new operator $n^{\dag}_{k}$ of $SU^{\ast}(3)$
will transform under the conjugate $SU(3)$ representation of
$(1,0)$, namely the IR $(0,1)$. Thus the allowed $SU^{\ast}(3)$
representations are given by
\begin{equation}
(\lambda,\mu)= \sum_{k=0}^{min(N_{p},N_{n})}
(N_{p}-k,N_{p}-k),\label{SU*3IRs}
\end{equation}
This corresponds to the reduction
\begin{align}
Sp(12,R) &\supset U(3,3)  \notag\\
&\supset SU_{p}(3) \otimes SU_{n}(3) \supset SU^{\ast}(3) \supset
SO(3), \label{NDS2}
\end{align}
i.e. through the maximal compact subalgebra $SU_{p}(3) \otimes
SU_{n}(3) \supset U(3,3)$. Consider for example $N_{p}=2$ and
$N_{n}=2$; then according to Eq. (\ref{SU*3IRs}) one finds
\begin{equation}
(2,0) \otimes (0,2) = (2,2)+(1,1)+(0,0).
\end{equation}
The $(2,2)$ irrep contains a $K=0$ band with $L=0,2$ and a $K=2$
band with $L=2,3,4$, while the $(1,1)$ irrep contains a $K=1$ band
with $L=1,2$ and the $(0,0)$ irrep contains a $K=0$ band with only
$L=0$. It is clear that this spectrum is very different from the
spectrum of $N_{p}=2$ and $N_{n}=2$ in the $SU(3)$ case. While in
the particle-particle case the ground band belongs to the
$(N_{p}+N_{p},0)$ irrep, in the particle-hole case the ground band
turns out to belong to the $(N_{p},N_{p})$ irrep.


The transformation (\ref{Aut}) corresponds to application of the
transformation $Q^{n}_{M} \rightarrow -Q^{n}_{M}$, $L^{n}_{M}
\rightarrow L^{n}_{M}$ in the n-boson $SU_{n}(3)$ algebra. This
changes the common $SU(3)$ quadrupole operator $Q_M =
Q^{p}_{M}+Q^{n}_{M}$ of the combined $pn$-system into that given by
(\ref{Qminus}). A similar type of $SU^{*}(3)$ algebra for IBM-2,
generated by $Q_M = Q^{p}_{M}-Q^{n}_{M}$ together with the angular
momentum operators is given in Ref.\cite{SU3AST2}. The
transformation (\ref{Aut}), being a special case of a wider class of
transformations known as inner automorphisms, does not change the
commutation relations of $SU(3)$ algebra, but however changes the
commutation relations of its complimentary $SU_{T}(2)$ algebra to
those corresponding to the non-compact subalgebra $SU(1,1) \subset
Sp(12,R)$ (see Eq.(\ref{chains})).

It is known that representation theory does provide all of the
embeddings, but it does not provide all of the dynamical symmetries
\cite{HS}. The inner automorphisms can provide new dynamical
symmetry limits, sometimes referred as to "hidden" \cite{HS} or
"parameter" symmetries \cite{parsym}. It will be shown in the next
sections that the (perturbed) $SU^{*}(3)$ algebra provides a new
physically distinct dynamical symmetry limit of the IVBM. Indeed,
the geometrical interpretation of this dynamical symmetry is that of
a prolate (proton) axially deformed rotor coupled to the oblate
(neutron) axially deformed rotor (or vice versa, when the inner
automorphism (\ref{Aut}) is performed with respect to the
$p-$bosons), which in some circumstances corresponds to a triaxial
shape of the compound nucleus in its ground-state configuration.

The most general Hamiltonian with $SU^{\ast}(3)$ symmetry consists
of the Casimir invariants of $SU^{\ast}(3)$ and its subgroup $SO(3)$
\begin{equation}
H=aC_{2}[SU^{\ast}(3)]+bC_{2}[SO(3)], \label{HSU*3}
\end{equation}
where
\begin{equation}
C_{2}[SU^{\ast}(3)]=\tfrac{1}{6}Q^2 +\tfrac{1}{2}L^{2}
\end{equation}
and the quadrupole operator Q is given by Eq. (\ref{Qminus}).

The spectrum of this Hamiltonian is determined by
\begin{equation}
H=a(\lambda^{2}+\mu^{2}+\lambda\mu+3\lambda+3\mu)+bL(L+1).
\label{ESU*3}
\end{equation}

\section{Shape structure}

In the present paper we are interested in the shapes corresponding
to the new dynamical symmetry limit. The geometry associated with a
given Hamiltonian can be obtained by the coherent state method. The
standard approach to obtain the geometry of the system is to express
the collective variables in the intrinsic (body-fixed) frame of
reference.

Within the IVBM, the (unnormalized) coherent state (CS) (or
intrinsic state) for the ground state band for even-even nuclei can
be expressed as \cite{PSIVBM}:
\begin{eqnarray}
\mid N;\xi,\zeta \ \rangle \propto \left[\sum_{k}
(\xi_{k}p_{k}^{\dagger}+\zeta_{k}n_{k}^{\dagger})\right]^N \mid 0 \
\rangle, \label{IVBMCS}
\end{eqnarray}
where the collective variables $\xi_{k}$ and $\zeta_{k}$ are
components of three-dimensional complex vectors. For static problems
these variables can be chosen real.

Usually, when some geometrical considerations concerning the choice
of the intrinsic frame are taken into account, the treatment of the
problem is significantly simplified. The geometry can be chosen such
that $\overrightarrow{\xi}$ and $\overrightarrow{\zeta}$  to span
the $xz$ plane with the $x$-axis along $\overrightarrow{\xi}$ and
$\overrightarrow{\zeta}$ is rotated by an angle $\theta$ about the
out-of-plane $y$-axis, $\overrightarrow{\xi} \cdot
\overrightarrow{\zeta}=r_{1}r_{2}\cos\theta$. In this way, the
condensate can be parametrized in terms of two real coordinates
$r_{1}$ and $r_{2}$ (the lengths of the two vectors), and their
relative angle $\theta$ ($r_{1},r_{2} \geq 0$  and $0 \leq \theta
\leq \pi$) \cite{PSIVBM}:
\begin{equation}
\mid N;r_{1},r_{2},\theta \ \rangle =
\frac{1}{\sqrt{N!}}(B^{\dagger})^{N}\mid 0 \ \rangle \label{BC2}
\end{equation}
with
\begin{equation}
B^{\dag}=\frac{1}{\sqrt{r_{1}^{2}+r_{2}^{2}}}\left[r_{1}
p^{\dagger}_{x} + r_{2}(n^{\dagger}_{x} \cos\theta +
n^{\dagger}_{z}\sin\theta)\right], \label{BdagDec}
\end{equation}
where $\mid 0 \ \rangle$ is the boson vacuum.

We simply study the $SU^{\ast}(3)$ Hamiltonian
\begin{equation}
H=kC_{2}[SU^{\ast}(3)]. \label{HSU*3p}
\end{equation}
expressed only by the second order $SU^{\ast}(3)$ Casimir operator.
In the present section, we set $k=-1$.

The ground-state energy surface is obtained by calculating the
expectation value of the boson Hamiltonian (\ref{HSU*3p}) in the CS
(\ref{BC2}):
\begin{equation}
E(N;r_{1},r_{2},\theta )=\frac{\langle
N;r_{1},r_{2},\theta|H|N;r_{1}r_{2},\theta \rangle}{\langle
N;r_{1},r_{2},\theta|N;r_{1},r_{2},\theta \rangle}. \label{Ener}
\end{equation}
The equilibrium "shape" is determined by minimizing the energy
surface with respect to $r_{1}$, $r_{2}$, and $\theta$. It is
convenient to introduce a new dynamical variable $\rho=r_{2}/r_{1}$
\cite{PSIVBM} as a measure of "deformation", which together with the
parameter $\theta$ determines the corresponding "shape".

\begin{figure}[h!]\centering
\includegraphics[width=80mm]{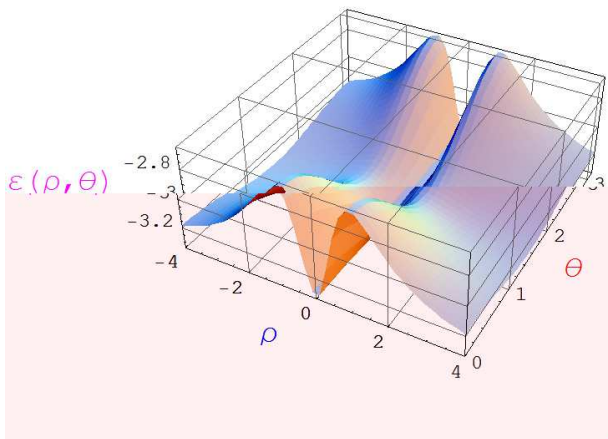}
\caption{(Color online) The scaled energy surface
$\varepsilon(\rho,\theta)$ in the $SU^{\ast}(3)$ limit for $k =
-1$.} \label{ESsu*3}
\end{figure}

The expectation value of Eq. (\ref{HSU*3p}) with respect to Eq.
(\ref{BC2}) gives the following energy surface
\begin{equation}
E(N;\rho,\theta)=\frac{2}{3}(kN)\left[\frac{1
+\rho^{4}-\rho^{2}(3\cos^{2}\theta-1)}{(1 +\rho^{2})^{2}} +4\right].
\label{ESSU*3}
\end{equation}
The scaled energy $\varepsilon(\rho,\theta)=E(N;\rho,\theta)/kN$ is
given in Figure \ref{ESsu*3}.

From the figure one can see that the global minimum occurs at
$\rho_{0} = 0$, which as will be shown further corresponds to an
oblate deformed shape.

In order to see to what geometry the energy surface (\ref{ESSU*3})
depicted in Fig. \ref{ESsu*3} corresponds, we consider the relation
of the parameter $\theta$ with the commonly used asymmetry parameter
$\gamma$ of the geometric CM of Bohr and Mottelson. A relation
between standard CM shape variables used to describe the deformation
of the collective motion and the shape parameters in the intrinsic
state of the IVBM can be obtained by calculating the expectation
value of the quadrupole moments of the corresponding dynamical
symmetry with respect to the IVBM coherent state. In the CS of the
IVBM, the effective $\gamma_{eff}$ deformation can be defined in the
usual way as \cite{BM}:
\begin{equation}
\tan \gamma_{eff} =\sqrt{2}\frac{\langle Q_{2} \rangle}{\langle
Q_{0} \rangle}, \label{Gamma}
\end{equation}
where $\langle Q_{\mu} \rangle$ denotes the expectation value of the
$\mu$th component of the quadrupole operator.

For the $SU^{\ast}(3)$ algebra with the generators (\ref{Qminus})
one obtains:
\begin{equation} \tan \gamma_{eff} =
\frac{\sqrt{3}(1-\rho^{2}\cos^{2}\theta)}{\left[-1-\frac{\rho^{2}}{2}(-3\cos2\theta+1)\right]}
. \label{GammaTita1}
\end{equation}

Expression (\ref{GammaTita1}) gives a relation between the
"projective" IVBM CS deformation parameters $\{\rho,\theta\}$ and
the standard collective model parameter $\gamma_{eff}$, determining
the triaxiality of the nuclear system. From Eq. (\ref{GammaTita1})
it is easily seen that for the equilibrium values of the IVBM shape
parameters $|\rho_{0}|=0$ and $\theta-$arbitrary in the
$SU^{\ast}(3)$ limit one obtains $|\gamma_{eff}|=60^{0}$ and hence
it corresponds to an oblate deformed shape.

Finally, we note that for $k > 0$ the minima of the $SU^{\ast}(3)$
energy surface (related to the maxima of Fig. \ref{ESsu*3} simply by
an inversion) are at $|\rho| \neq 0$ ($|\rho| = 1$) and
$\theta_{0}=0^{0}$. For $k > 0$, there is a second (local) extremum
placed at $|\rho_{0}|=1$ and $\theta_{0}=90^{0}$, which according to
Eq. (\ref{GammaTita1}) corresponds to $|\gamma_{eff}|=30^{0}$ and
hence to a triaxial maximum. In the next section we will see that
the addition of some perturbation terms to the Hamiltonian
(\ref{HSU*3p}) changes the structure of the energy surface and a
stable triaxial minimum appears.

\section{Perturbation of the SU*(3) dynamical symmetry}

As it was mentioned, the exact $SU^{\ast}(3)$ symmetry shows a large
degeneracy in the level spectra which in actual nuclei is not
observed. In some cases, the energy spectra can be situated in
between the $SU^{\ast}(3)$ and $O(6)$ or $SU(3)$ and $SU^{\ast}(3)$
dynamical limits. Indeed, several systematic studies \cite{SS} have
shown that transitional nuclei exhibit the triaxial features. A
number of signatures of $\gamma-$soft and $\gamma-$rigid structures
in nuclei has been discussed \cite{exp1},\cite{exp2},\cite{SS}. In
Ref. \cite{O6tr} it was shown that the empirical deviations from the
$O(6)$ limit of the IBM, in the Pt and Xe, Ba regions, can be
interpreted by introducing explicitly triaxial degrees of freedom,
suggesting a more complex and possibly intermediate situation
between $\gamma-$rigid and $\gamma-$unstable properties. In this
respect, we study the influence of different types of perturbations
on the $SU^{\ast}(3)$ dynamical symmetry of the IVBM. We consider
only the two types of perturbation terms on the $SU^{\ast}(3)$
energy surface, namely the inclusion of a Majorana interaction and
an $O(6)$ term.

\subsection{The Majorana perturbation}

The Hamiltonian, to which a Majorana term is added, takes the form
\begin{equation}
H_{I}= k\frac{1}{N-1}C_{2}[SU^{\ast}(3)] + a\frac{1}{N-1}M_{3},
\label{HpM3}
\end{equation}
where the Majorana operator is defined as
\begin{equation}
M_{3}=2(p^{\dag} \times n^{\dag})^{(1)} \cdot (p \times n)^{(1)}
\label{M3}
\end{equation}
and it is related to the $U(3)$ second order Casimir invariant
$C_{2}[U(3)]$ via the relation
\begin{equation}
C_{2}[U(3)]=N(N+2)-2M_{3}. \label{M3C2}
\end{equation}
In Eq. (\ref{HpM3}) the appropriate scaling factors in $N$ are
included.

\begin{figure}[h]\centering
\includegraphics[width=80mm]{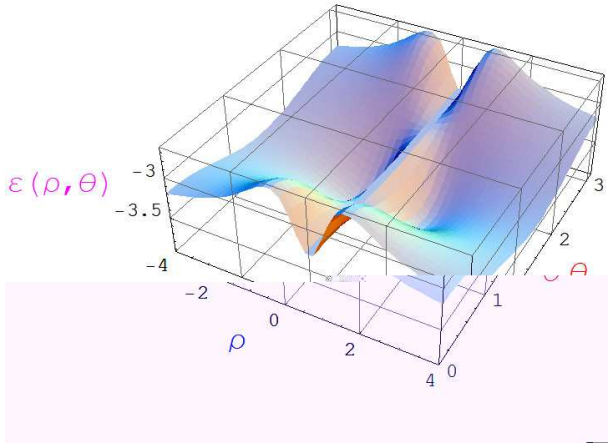}
\caption{(Color online) The scaled energy surface
$\varepsilon(\rho,\theta)$ in the $SU^{\ast}(3)$ limit, when a
Majorana term is added. The values of the model parameters used are
$k = -1$, $a = -3$.} \label{Maj1}
\end{figure}

\begin{figure}[h]\centering
\includegraphics[width=60mm]{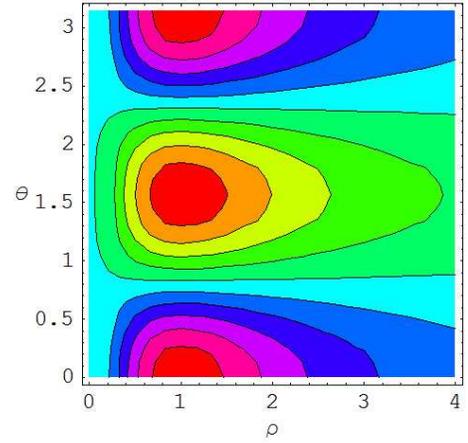}
\caption{(Color online) Contour plot of the scaled energy surface
$\varepsilon(\rho,\theta)$ in the $SU^{\ast}(3)$ limit, when a
Majorana term is added. The values of the model parameters used are
$k = -1$, $a = -3$. Only the region $\rho > 0$ is depicted.}
\label{Maj2}
\end{figure}

The classical limit of the Majorana term (\ref{M3}) is given by the
following expression
\begin{equation}
E(N;\rho,\theta)=aN(N-1)\frac{\rho^{2}\sin^{2}\theta}{(1+\rho^{2})^{2}},
\label{ESM3}
\end{equation}
The scaled energy surface of the Hamiltonian (\ref{HpM3}) (Eqs.
(\ref{ESSU*3}) and (\ref{ESM3})) is shown in Figures \ref{Maj1} and
\ref{Maj2} in the form of a three-dimensional plot and a contour
plot, respectively. The values of the model parameters used are $k =
-1$, $a = -3$. From the figures, according to Eq.(\ref{GammaTita1}),
it is clear that a stable triaxial minimum results at
$\theta_{0}=90^{0}$ and $|\rho_{0}|=1$, which becomes deeper and
deeper with the increasing of the absolute value of the parameter
$a$.

The inspection of the energy surfaces for different values of the
parameter $a$ (at fixed $k = -1$) shows that for small negative
values of the parameter $a$ ($|a| \leq 0.4$) and realistic values of
$\rho \in [0,1.5]$, the minimum is at $\rho_{0}=0$, corresponding to
an oblate deformed shape. In the interval $|a| \approx 0.5-0.8$
there exist two degenerate minima at $|\rho_{0}|=0$ and
$|\rho_{0}|=1$, $\theta_{0}=90^{0}$ respectively, while for $a \leq
- 0.86$ a stable triaxial minimum ($\theta_{0}=90^{0}$,
$|\rho_{0}|=1$) occurs. This triaxial minimum persists and for
positive values of the parameter $k$ ($k = 1$) when $a < -2$ and
also becomes more pronounced with the further increasing of the
absolute value of $a$.

\begin{figure}[h]\centering
\includegraphics[width=100mm]{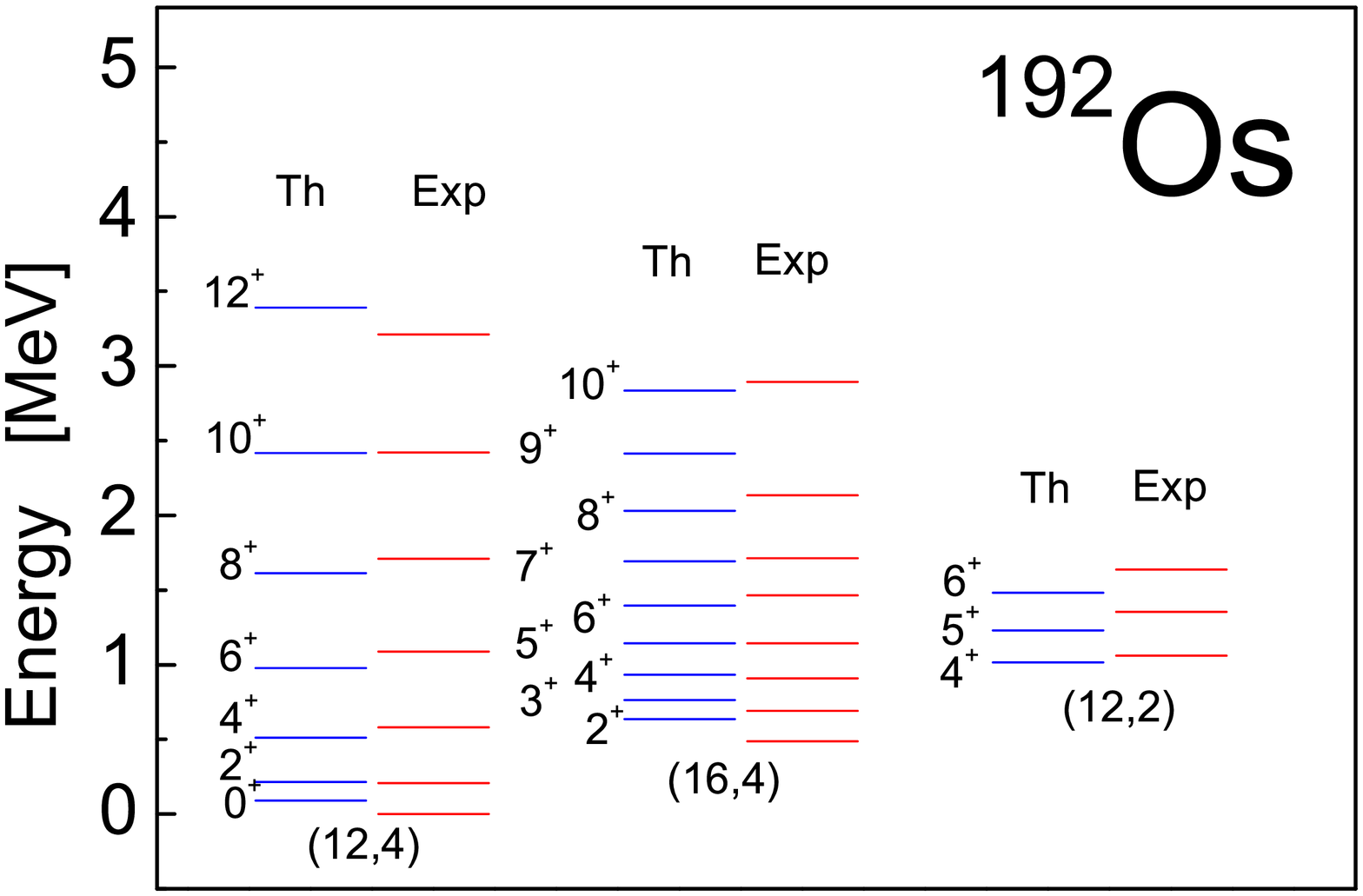}
\caption{(Color online) Detailed energy spectrum, obtained with the
Hamiltonian (\ref{HpM3LL}), corresponding to the parameters $k =
-0.0136$ MeV, $k' =0.0343$ MeV and $a =-0.0131$ MeV, compared with
the experimental data for $^{192}$Os. Different $SU^{\ast}(3)$
irreps associated with the bands under consideration are also
indicated. Data are taken from Refs.\cite{SP},\cite{os192exp}.}
\label{energy}
\end{figure}

In the present work we are mainly interested in the ground state
properties of the energy surfaces considered. Nevertheless, in order
to see to what extent the structure of the theoretical energy levels
corresponds to a real experimental pattern and how the energy
spectrum generated by the pure $SU^{\ast}(3)$ Hamiltonian
(\ref{HSU*3p}) is influenced by the inclusion of the Majorana
interaction, we consider the following Hamiltonian
\begin{equation}
H=kC_{2}[SU^{\ast}(3)]  + k'C_{2}[SO(3)] + aM_{3}, \label{HpM3LL}
\end{equation}
where the rotational term in Eq.(\ref{HpM3LL}) is added to lift the
degeneracy of the states with different angular momentum.


\begin{figure}[h]\centering
\includegraphics[width=80mm]{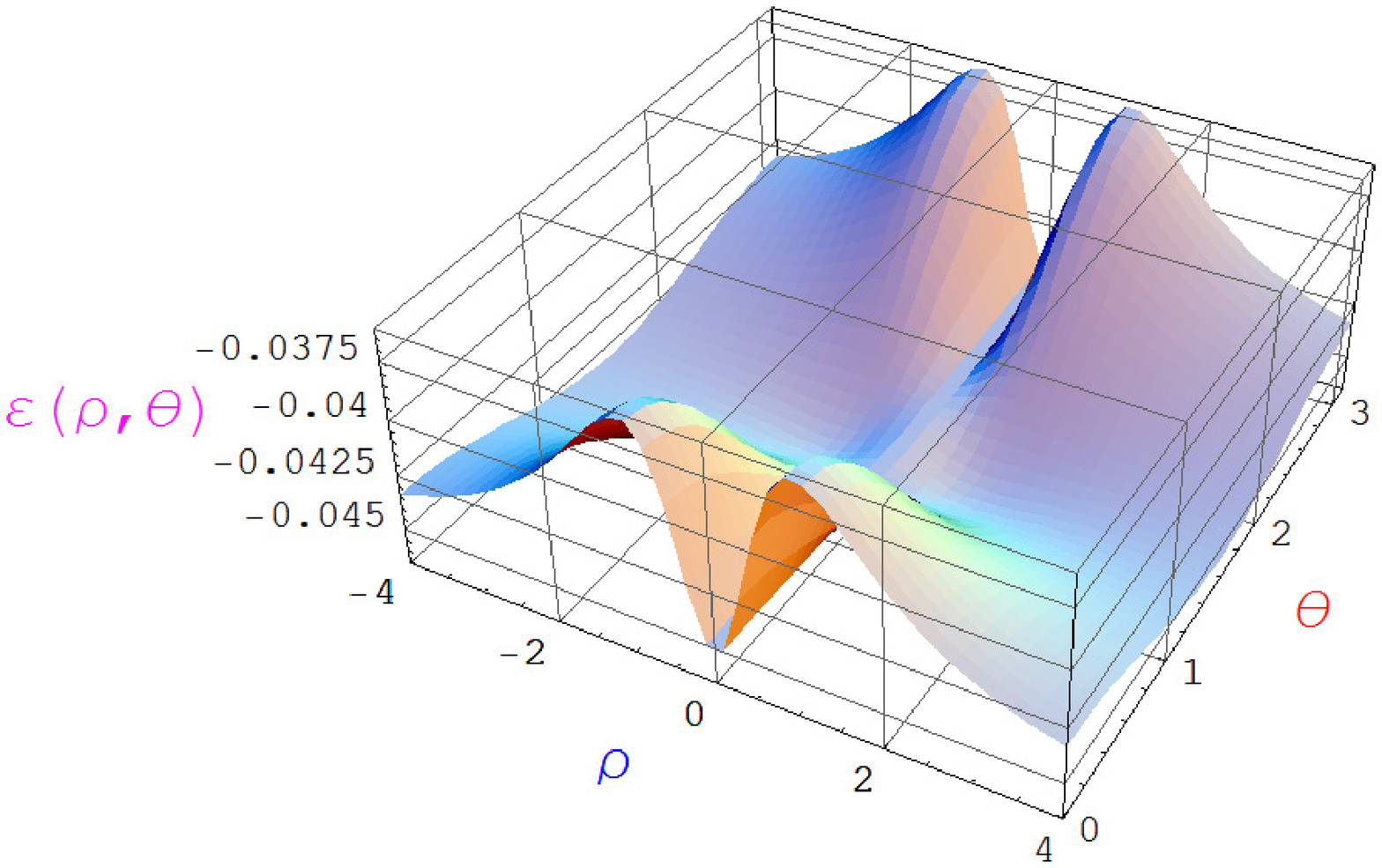}\hspace{20.mm}
\includegraphics[width=60mm]{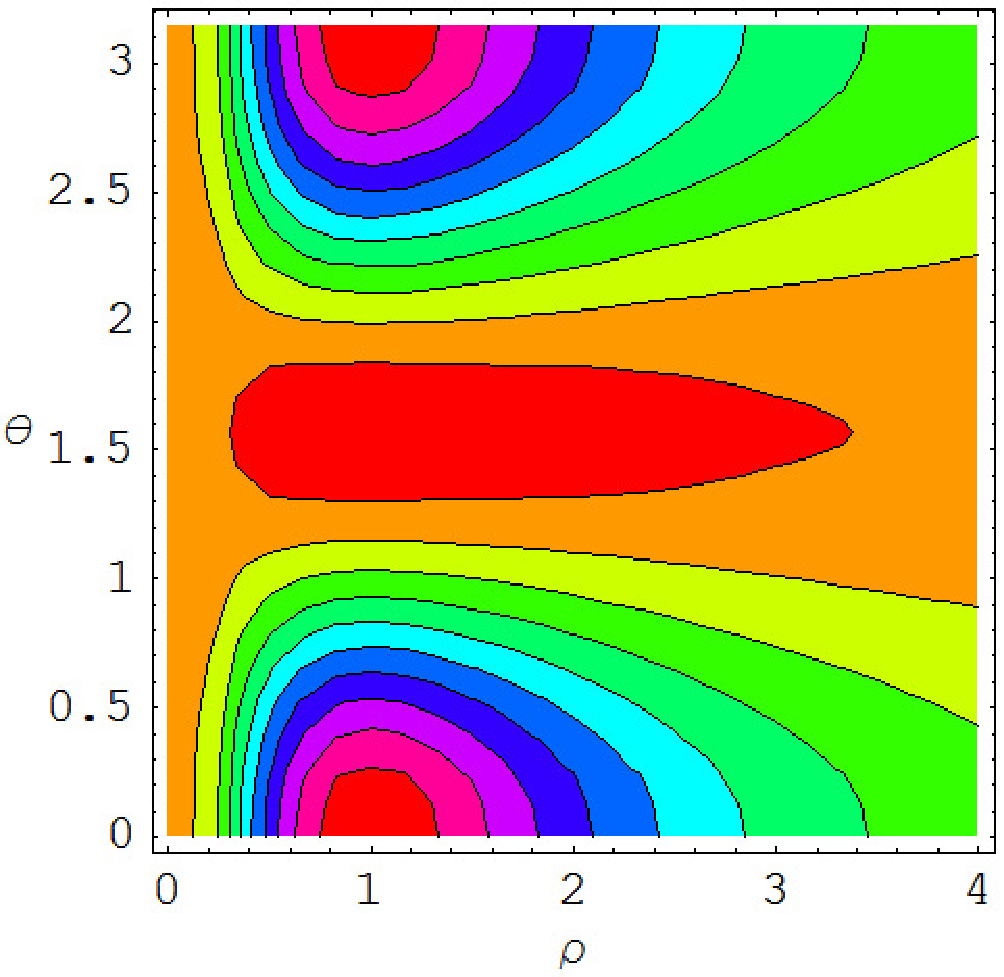}
\caption{(Color online) The scaled ground state energy surface
$\varepsilon(\rho,\theta)$ in $^{192}$Os for the model parameters
obtained in the fitting procedure in the form of three-dimensional
and contour plots.} \label{Os192ES}
\end{figure}


In Fig. \ref{energy} we plot the theoretical predictions for the
ground state band, $\gamma$ band and $K^{+}=4^{+}$ band energies,
obtained with the Hamiltonian (\ref{HpM3LL}) with the following
values of the model parameters $k = -0.0136$ MeV, $k' =0.0343$  MeV
and $a =-0.0131$  MeV. The values of the latter are determined by
using a minimization $\chi^{2}$-procedure. The theoretical results
are compared to the experimental data \cite{SP},\cite{os192exp} for
the nucleus $^{192}$Os. The latter is considered in the literature
(see, e.g., Refs.\cite{exp2},\cite{SP}) as being a triaxial one.
From the figure one can see that the theoretical predictions are far
from perfect (especially for the GSB), but nevertheless the
structure of the energy spectrum of $^{192}$Os is reasonably
reproduced in general. The fit is performed for all states of the
ground state band and $\gamma$ band simultaneously. That is why we
obtain an average fit along the whole bands and the states at the
bottom for these two bands are overestimated.

The quality of the obtained results is not surprising taking into
account the very simple form of the Hamiltonian which is used. The
improvement of the theoretical results obviously requires a more
realistic interaction which should be incorporated into the model
Hamiltonian. As we have said, the present work is focused on the
ground state properties of the energy surface and the calculations
carry a very schematic character.

We plot the ground state energy surface in $^{192}$Os for the model
parameters obtained in the fitting procedure in the form of
three-dimensional and contour plots in Fig. \ref{Os192ES}. From the
figure one can see that a very shallow triaxial minimum for the
ground state in $^{192}$Os is observed, which corresponds to
$\gamma_{eff}=30^{0}$.  The latter is separated from the neighboring
oblate minimum by only $\simeq 1$ keV (see the energy scale in Fig.
\ref{Os192ES}), i.e. the two observed minima are practically
degenerate. From the contour plot in Fig. \ref{Os192ES} it can be
seen that this extremely shallow triaxial minimum is soft along the
$\theta$ direction, which corresponds to $\gamma$-softness (the
change of $\rho$ at fixed $\theta_{0}=90^{0}$ changes the asymmetry
parameter $\gamma_{eff}$). The structure of the energy surface
obtained in our schematic calculations for $^{192}$Os supports the
consideration of this nucleus as being a transitional one between
axially symmetric prolate and oblate deformed ones, passing through
a $\gamma$-soft triaxial region. Indeed, some theoretical
calculations \cite{CCQH}, \cite{Robledo} predict a very tiny region
of triaxiality between the prolate and oblate shapes. The
self-consistent Hartree-Fock-Bogoliubov calculations with Gogny D1S
and Skyrme SLy4 forces predict that the prolate to oblate transition
takes place at neutron number $N=116$, i.e. exactly the case for
$^{192}$Os.

\begin{figure}[h]\centering
\includegraphics[width=75mm]{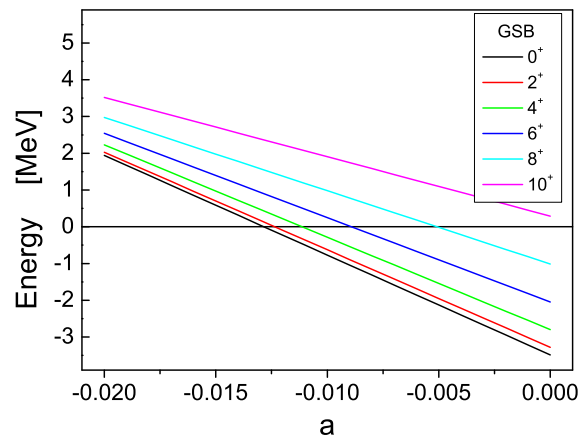}
\caption{(Color online) The ground state band for the Hamiltonian
(\ref{HpM3LL}) as a function of the strength parameter $a$. The
values of the rest model parameters are $k = -0.0136$ MeV, $k'
=0.0343$ MeV. The $SU^{\ast}(3)$ irrep corresponding to the GSB is
$(12,4)$.} \label{EHpMLL-GSB}
\end{figure}

\begin{figure}[h]\centering
\includegraphics[width=75mm]{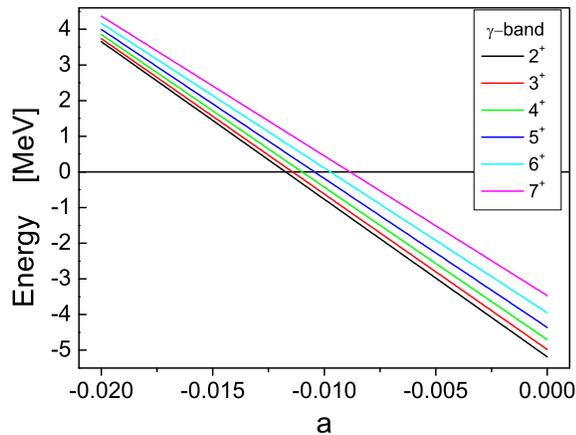}
\caption{(Color online) The $\gamma$ band for the Hamiltonian
(\ref{HpM3LL}) as a function of the strength parameter $a$. The
values of the rest model parameters are $k = -0.0136$ MeV, $k'
=0.0343$ MeV. The $SU^{\ast}(3)$ irrep corresponding to the $\gamma$
band is $(10,6)$.} \label{EHpMLL-GAMA}
\end{figure}

The evolution of the ground state band and $\gamma$ band for the
Hamiltonian (\ref{HpM3LL}) as a function of the strength parameter
$a$ is shown in Figs. \ref{EHpMLL-GSB} and  \ref{EHpMLL-GAMA},
respectively. The values of the rest model parameters are kept the
same as given above. From the figures one can see that the inclusion
of the Majorana term does not change the level spacings for both the
ground state and $\gamma$ bands and hence preserves the character of
the bands. One can see also that the energy levels of both the GSB
and the $\gamma$ band are affected in the same manner as a function
of the strength of the Majorana interaction.

\subsection{Phase transition between O(6) and SU*(3) limits}

The transition from the $SU^{\ast}(3)$ to the $O(6)$ limit can be
realized by the following Hamiltonian
\begin{equation}
H_{II}=(1-g)\frac{1}{N-1}P^{\dag}P  - g\frac{1}{N-1}
C_{2}[SU^{\ast}(3)], \label{Hmix}
\end{equation}
varying the parameter $g$ from $g=0$ ($O(6)-\gamma$-unstable limit)
to $g=1$ ($SU^{\ast}(3)-$ limit). The $O(6)$ pairing operator is
defined as $P^{\dag}=\frac{1}{2}(p^{\dag} \cdot p^{\dag}-n^{\dag}
\cdot n^{\dag})$. The $P^{\dag}P$ operator in Eq. (\ref{Hmix}) is
related to the quadratic Casimir operator $C_{2}[O(6)]$ of $O(6)$ by
the equation $C_{2}[O(6)]=-4P^{\dag}P + N(N+4)$.

In Figure \ref{mix1} we show the scaled energy surfaces
corresponding to the Hamiltonian (\ref{Hmix}) for three different
values of g, namely $g = 0,0.4$ and $0.65$, respectively. For $g =
0$ we have the typical energy surface for the $\gamma$-unstable
deformed shape.

The evolution of the energy surfaces for the same values of the
parameter $g$ is shown as contour plots in Figure \ref{mix1c}.
Numerical studies show that the triaxial minimum
($\theta_{0}=90^{0}$,$|\rho_{0}|=1$) persists for the values of $g$
in the interval $0 < g \leq 0.85$, where for small values of $g$ it
is very shallow and becomes more pronounced with the increase of $g$
(up to $g \approx 0.8$). From Fig.\ref{mix1} it can be seen that a
second local minimum appears at $\rho_{0} = 0$  for $g = 0.65$.
Around $g \sim 0.84$ the two minima become degenerate (up to $g
\simeq 0.88$) and for $g \geq 0.89$ the second minimum at $\rho_{0}
= 0$ becomes a global one in the interval $\rho \in [0,1.5]$ (just
as in the case when the $SU^{\ast}(3)$ symmetry is perturbed by the
Majorana interaction for comparatively small values of the parameter
$a$, $|a|<0.7$). The geometry of the $SU^{\ast}(3)$ ($g=1$) limit
for $\rho_{0}=0$, as was mentioned, corresponds to that of an oblate
deformed rotor ($|\gamma_{eff}|=60^{0}$).

From the results obtained in the last two sections it can be
concluded that the two types of perturbations disturb the exact
$SU^{\ast}(3)$ symmetry energy surface in a similar way and lead to
the same geometrical structure underlying both Hamiltonians under
consideration. \\

\begin{widetext}

\begin{figure}[h]\centering
\includegraphics[width=58mm]{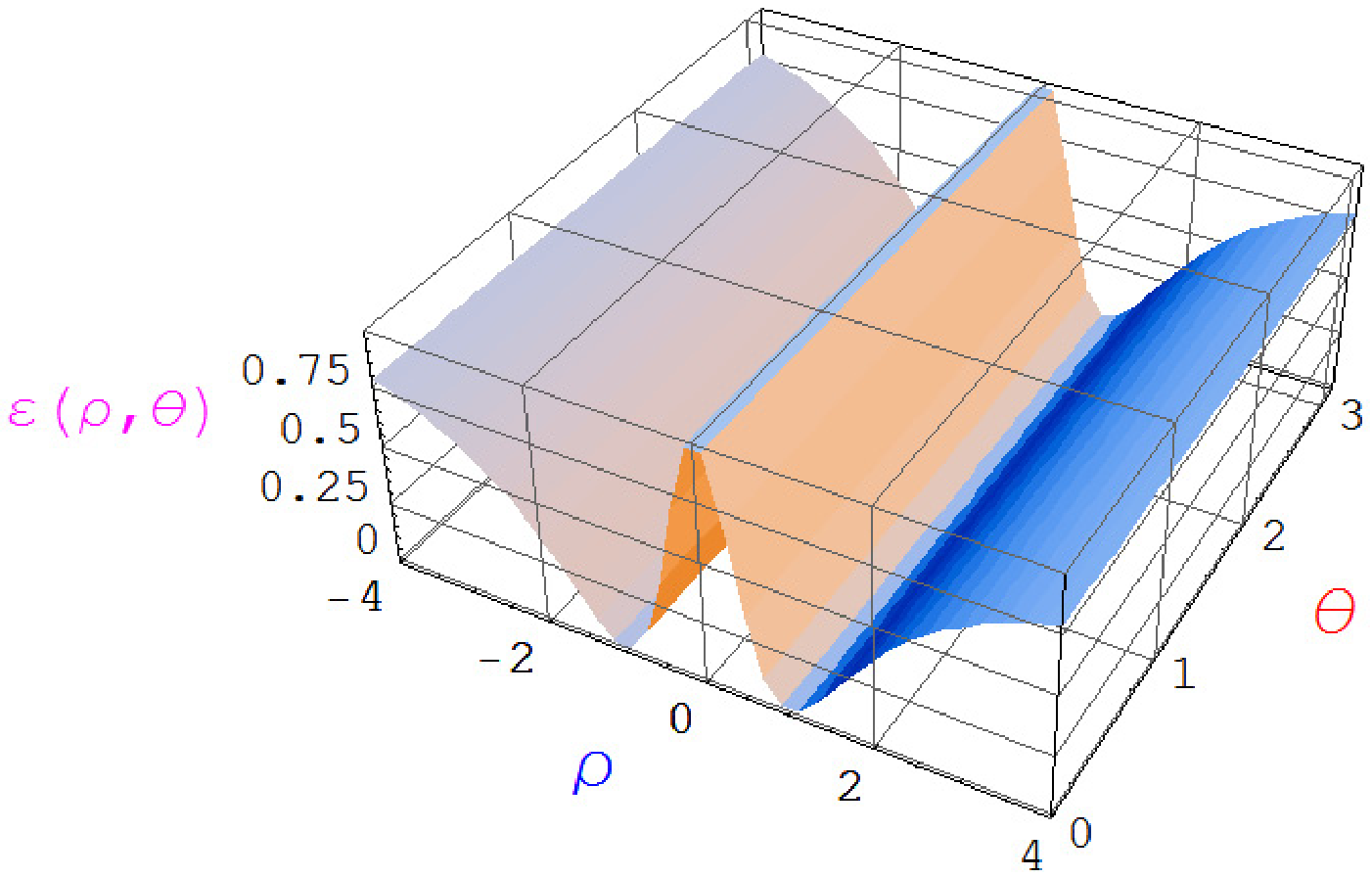}\hspace{1.mm}
\includegraphics[width=58mm]{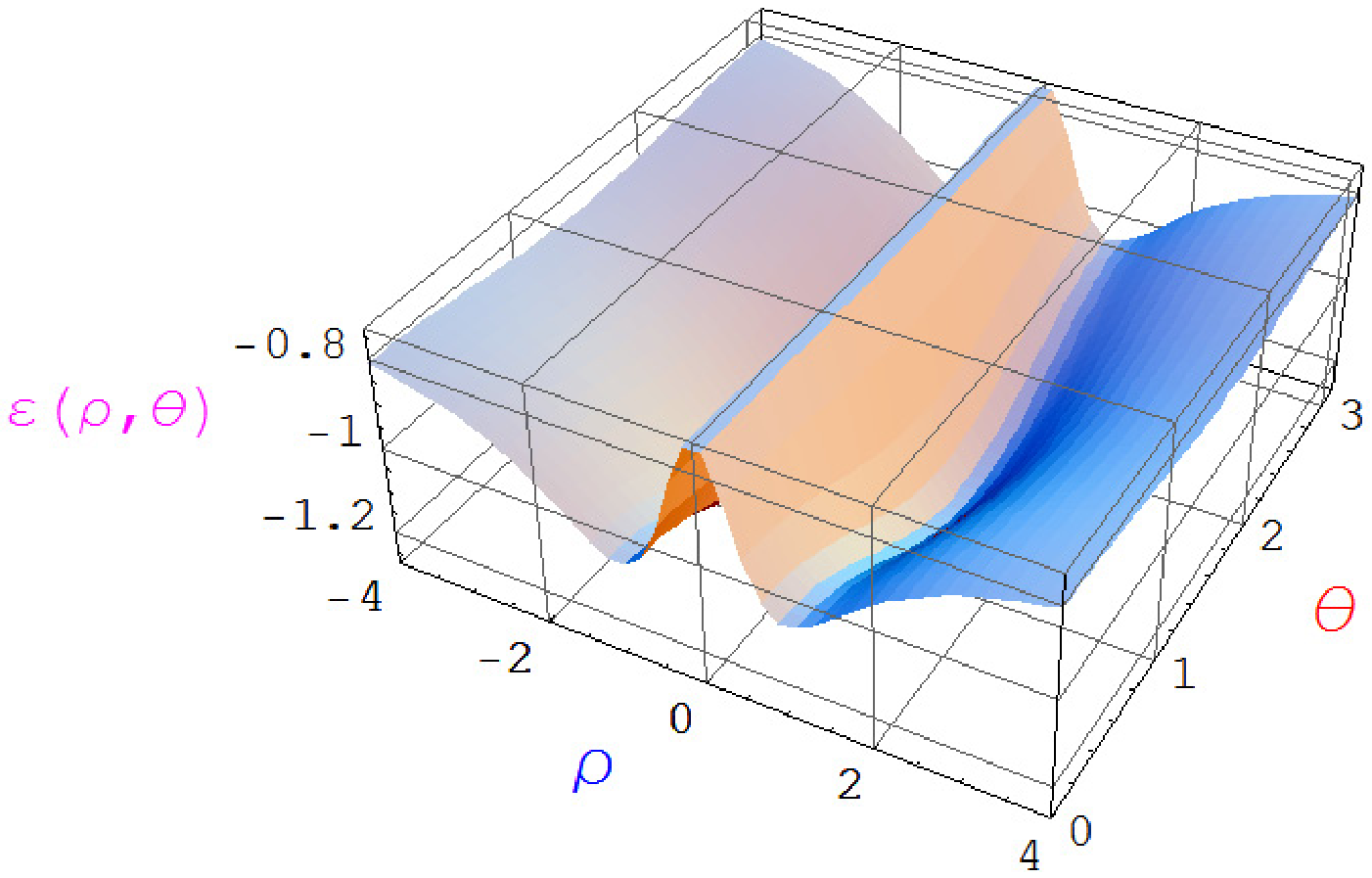}\hspace{1.mm}
\includegraphics[width=58mm]{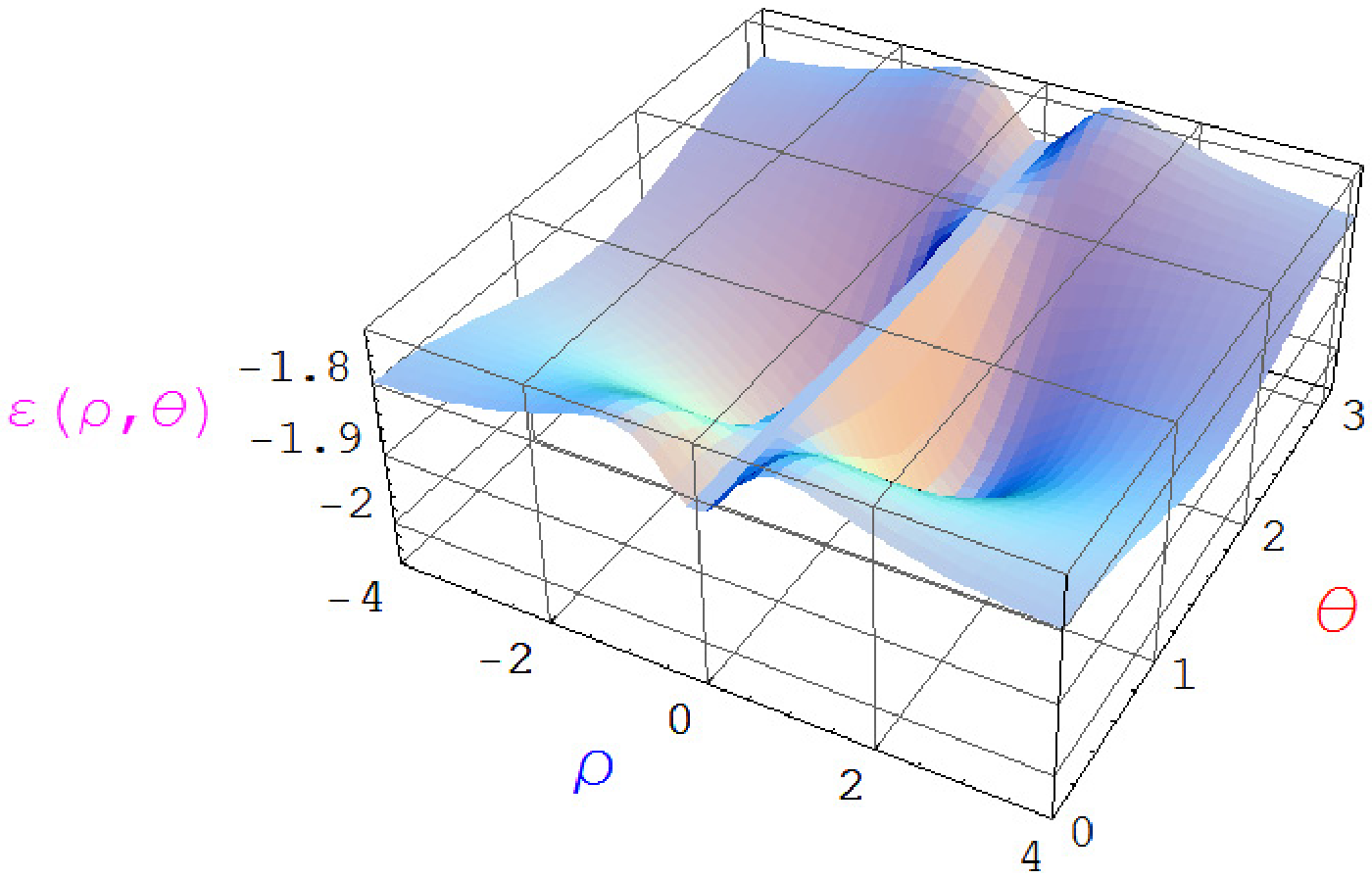}
\caption{(Color online) The scaled energy surface
$\varepsilon(\rho,\theta)$ corresponding to the Hamiltonian
(\ref{Hmix})  for $g = 0, 0.4$ and 0.65, respectively.} \label{mix1}
\end{figure}

\begin{figure}[h]\centering
\includegraphics[width=50mm]{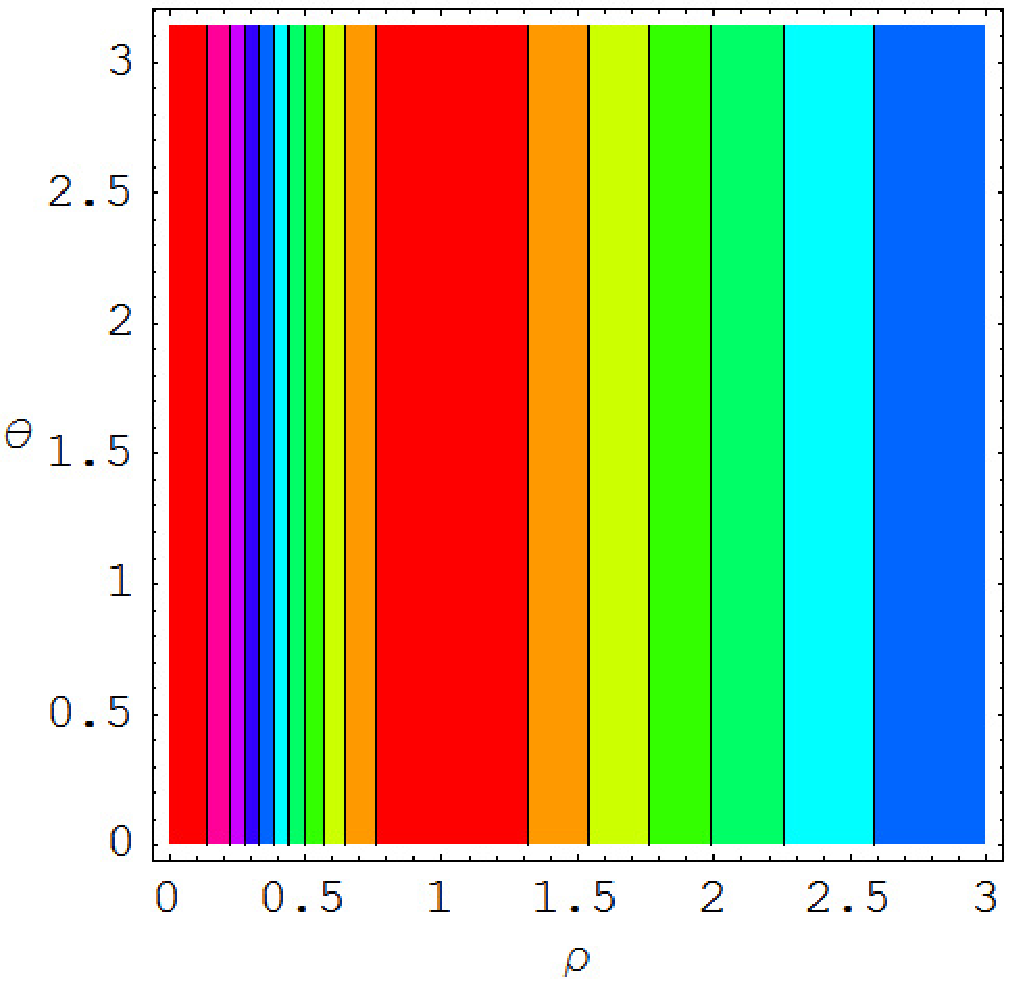}\hspace{1.mm}
\includegraphics[width=50mm]{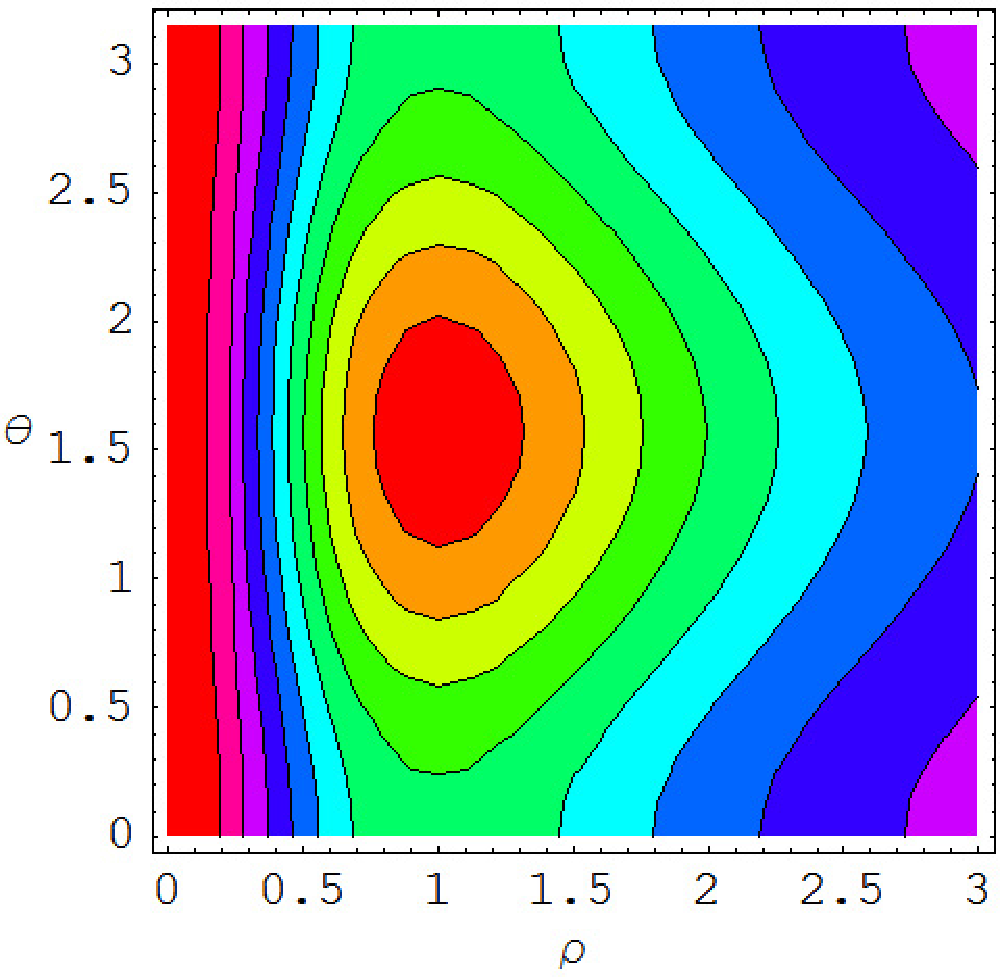}\hspace{1.mm}
\includegraphics[width=50mm]{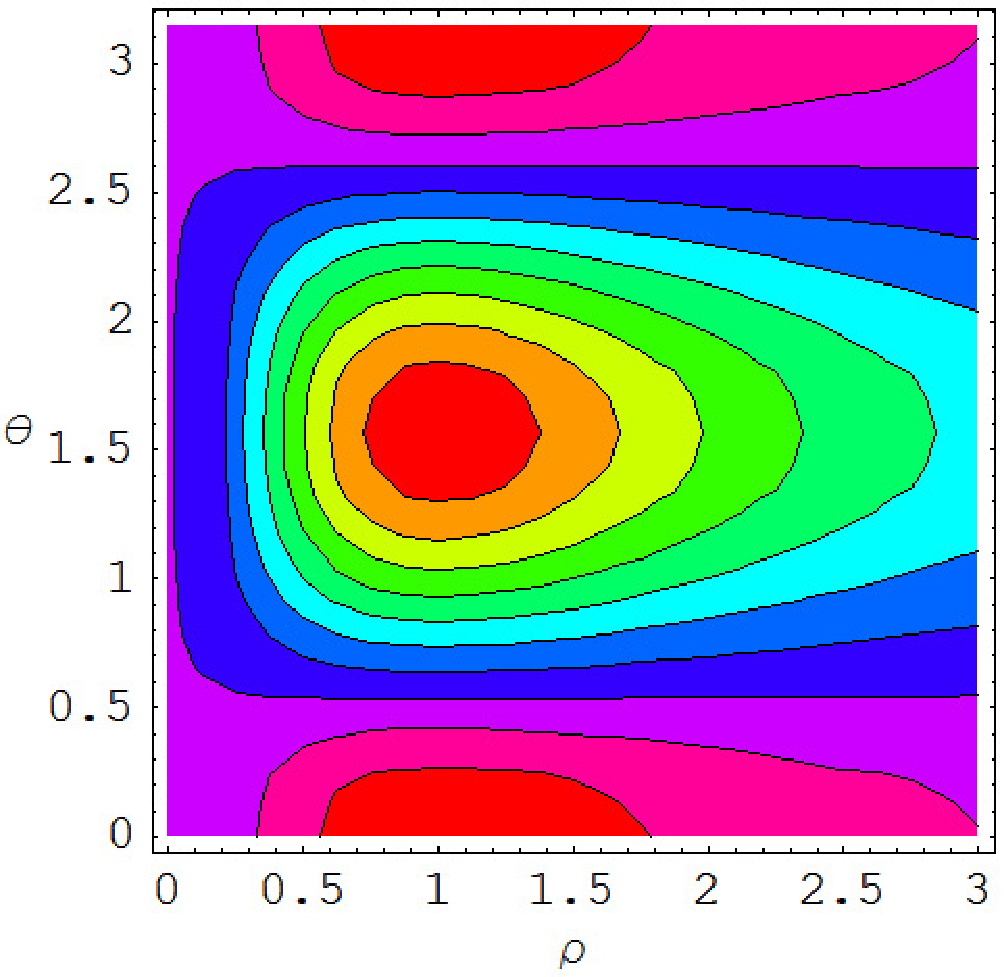}
\caption{(Color online) A contour plot of the scaled energy surface
$\varepsilon(\rho,\theta)$ corresponding to the Hamiltonian
(\ref{Hmix}) for $g = 0, 0.4$ and 0.65, respectively. Only the
region $\rho > 0$ is depicted.} \label{mix1c}
\end{figure}

\end{widetext}

\section{Summary}

A new dynamical symmetry limit of the two-fluid Interacting Vector
Boson Model, defined through the chain $Sp(12,R) \supset U(3,3)
\supset U^{\ast}(3) \otimes SU(1,1) \supset SU^{\ast}(3) \supset
SO(3)$, is introduced. The $SU^{\ast}(3)$ algebra considered in the
present paper closely resembles many properties of the
$SU^{\ast}(3)$ limit of IBM-2, which have been shown by many authors
geometrically to correspond to the rigid triaxial model.

We have studied the influence of different types of perturbations on
the $SU^{\ast}(3)$ dynamical symmetry energy surface. In particular,
the addition of a Majorana interaction and an $O(6)$ term to the
model $SU^{\ast}(3)$ Hamiltonian is investigated. It is shown that
the effect of these perturbations results in the formation of a
stable triaxial minimum in the energy surface of the IVBM
Hamiltonian under consideration.

The effect of the Majorana interaction on the energy levels of the
ground state band and the $\gamma$-band is studied as well. Using a
schematic Hamiltonian (possessing a disturbed $SU^{\ast}(3)$
dynamical symmetry) the theory is applied for the calculation of the
low-lying energy spectrum of the nucleus $^{192}$Os, which has been
considered in the literature as being triaxial. The theoretical
results obtained agree reasonably with the experimental data and
show a very shallow triaxial minimum in the energy surface for the
ground state in $^{192}$Os. This suggests that the newly proposed
dynamical symmetry might be appropriate for the description of the
collective properties of different nuclei, exhibiting triaxial
features. More investigations in this direction are further
required.

\section*{Acknowledgment}

This work was supported by the Bulgarian National Foundation for
scientific research under Grant Number DID-$02/16/17.12.2009$.

\end{document}